\documentclass{osa-article}

\graphicspath{{./figures/}}

\usepackage{tabularx}
\usepackage{booktabs}
\usepackage{threeparttable}
\usepackage{graphicx}
\usepackage{subcaption}
\usepackage{lineno}
\usepackage{hyperref}
\hypersetup{
        colorlinks=true,
        citecolor=SteelBlue,
        filecolor=LimeGreen,
        linkcolor=SlateBlue,
        urlcolor=MediumPurple
}

\newcolumntype{Y}{>{\centering\arraybackslash}X}
\newcolumntype{s}{>{\centering\arraybackslash \hsize=.75\hsize}X}
\usepackage{cleveref}

\newcommand{\G}[1]{\textcolor{SteelBlue}{#1}} 

\journal{oe}

\articletype{Research Article}

\begin{document}

\title{Global optimization of multilayer dielectric coatings for precision
  measurements}

\author{Gautam Venugopalan,\authormark{1,*}, Francisco Salces-C\'arcoba,\authormark{1}, Koji Arai,\authormark{1}, Rana X Adhikari\authormark{1}  }

\address{\authormark{1}LIGO Laboratory, California Institute of Technology, MS 100-36, Pasadena, CA 91125, USA}

\email{\authormark{*}gautam@caltech.edu} 



\begin{abstract}
We describe the design of optimized multilayer dielectric coatings for precision laser interferometry.
By setting up an appropriate cost function and then using a global optimizer to find a minimum in the parameter space, we were able to realize coating designs that meet the design requirements for spectral reflectivity, thermal noise, absorption, and tolerances to coating fabrication errors. We also present application of a Markov-Chain Monte Carlo (MCMC) based parameter estimation algorithm that can infer thicknesses of dielectric layers in a coating, given a measurement of the spectral reflectivity. This technique can be a powerful diagnostic tool for both commercial coating manufacturers, and the community using dielectric mirrors for precision metrology experiments.
\end{abstract}

\section{Introduction}
\label{sec:intro}
Interferometric gravitational wave detectors such as Advanced LIGO require mirrors with dielectric coatings that satisfy multiple requirements on reflectivity at one or more wavelengths, surface electric field, absorption within the coating, and thermal noise~\cite{Martynov:16}. For the next generation of detectors, it is anticipated that there will be tighter requirements on these specifications. Furthermore, it is desirable that the coating design chosen will have minimal sensitivity to manufacturing tolerances.

In this paper, we describe the construction of a cost function that quantifies how closely a given coating design satisfies the multiple requirements on it. Once this cost function has been constructed, it can be minimized with a numerical optimization algorithm, such as Particle Swarm Optimization~\cite{MatlabPSO} or Differential Evolution \cite{2020SciPy-NMeth,Storn:97}, to compute the coating design (i.e. the set of thicknesses of various layers in a dielectric stack of alternating high- and low-index materials) that gives the (global) minimum value in the allowed parameter space. Since most numerical optimization algorithms are designed to handle a scalar cost function, we convert our vector (multi-objective) cost function to a scalar number by taking the scalar product of it with a weight vector. The latter allows us to quantify which design objectives are more important than others.

Once an optimal solution has been arrived at, we verify its sensitivity to small perturbations in layer thicknesses as well as various assumed model parameters using Monte-Carlo (MC) analysis. This approach allows us to evaluate the relative performance of coating designs that vary along `hyper-parameter' axes, such as the choice of number of layer pairs that make up the coating.

This approach of mapping a complex design problem into a scalar objective function minimization problem can be readily generalized to several other problems in the field of Gravitational Wave Astronomy. It is also of interest to the broader category of experiments that use optical cavities for precision measurements in the atomic and molecular optics community.

\section{Requirements on multilayer dielectric coatings}
\label{sec:coatingMath}
In this section, the requirements on dielectric coatings of the type considered in this paper are described. Although the method is easily generalized to a multilayer stack with arbitrary refractive index profile, we only consider cases in which the stack is comprised of alternating layers of two dielectrics, with refractive indices $n_1$ (low-index) and $n_2$ (high-index). The coating is deposited on a substrate with refractive index $n_{\mathrm{sub}}$. These considerations are typical in the field of laser interferometry, but are also of relevance in various other fields such as nanoscale optomechanics. Strategies to increase the numerical efficiency of computing these properties are also briefly discussed.

\subsection{Spectral reflectivity}\label{sec:2.1}
The primary requirement on a dielectric coating design is the \textit{power} reflectivity at wavelengths of interest, $R(\lambda)$. For a given dielectric coating stack with $M$ interfaces, this may be calculated in several ways. We opt to do the computation of the \textit{amplitude} reflectivity, $\Gamma$ recursively, using the relation \cite{Orfanidis:2016}

\begin{equation}
\Gamma_i = \frac{\rho_i + \Gamma_{i+1}e^{-2i k_i l_i}}{1 + \rho_i \Gamma_{i+1}e^{-2i k_i l_i}},
\label{eq:2.1}
\end{equation}

where the index $i = M, M-1, ..., 1$, with $i = 1$ corresponding to the interface of the coating with the incident electromagnetic field. In \Cref{eq:2.1},

\begin{subequations}
	\begin{equation}
		k_i = \frac{2 \pi n^T_i \mathrm{cos}\theta}{\lambda},
	\label{eq:2.2.1}
	\end{equation}
	\begin{equation}
		\rho_i = \frac{n^T_{i-1} - n^T_i}{n^T_{i-1} + n^T_i},
	\label{eq:2.2.2}
	\end{equation}
	\begin{equation}
		n^T_i = \begin{cases}
		\frac{n_i}{\mathrm{cos}\theta _i},  & \text{p-polarization},\\
		n_i \mathrm{cos}\theta _i, & \text{s-polarization},
		\end{cases}
	\label{eq:2.2.3}
	\end{equation}
\label{eq:2.2}
\end{subequations}
where $l_i$ is the \textit{physical} thickness of, $n_i$ is the (wavelength dependent) refractive index of, and $\theta _i$ is the angle of incidence into the $i-\text{th}$ layer. The recursion relation \Cref{eq:2.1} is initialized with $\Gamma_{M+1} = \rho_{M+1} = \frac{n^T_M - n^T_{\mathrm{sub}}}{n^T_M + n^T_{\mathrm{sub}}}$, where $n^T_{\mathrm{sub}}$ is defined by \Cref{eq:2.2.2} for the substrate onto which the dielectric stack is deposited. The power reflectivity of the stack may then be computed as $R = |\Gamma_1|^2$. A typical coating design requirement will specify $R(\lambda, \theta, \mathrm{polarization})$.

\subsection{Surface electric field and absorption}
In optical cavities in which the circulating power is high, absorption in the dielectric layers becomes important for a number of reasons.
The coatings have to be able to withstand the absorption-induced thermal heating for the highest expected incident electromagnetic field intensity.
In applications, where the cavity mirrors have to be maintained at cryogenic temperatures, the requirement becomes even more stringent as the rate at which heat can be extracted from the optic will set the maximum permissible absorption in the coating~\cite{Voyager:Science}.

Absorption is quoted as a dimensionless fraction of the incident power which is converted to thermal energy in the coating. We evaluate absorption by first determining the square of the electric field as a function of penetration depth, $\mathrm{z}$, normalized by the incident electric field, $N=\frac{|\vec{E}(\mathrm{z})|^2}{|\vec{E}_o^+|^2}$\cite{Arnon:80}. The total absorption, $\alpha_{\mathrm{T}}$ in a coating of thickness $L$ is
\begin{equation}
\alpha_{\mathrm{T}} = \int_{0}^{\mathrm{L}} d\mathrm{z} \, N \alpha (\mathrm{z}),
\label{eq:2.3}
\end{equation}

with $\alpha (\mathrm{z})$ describing the bulk absorption of the materials used in the coating, for which measured values are available. In practise, the integral in \Cref{eq:2.3} can be evaluated numerically. However, this is a computationally expensive operation. A good proxy for use in an optimizer is the value of the electric field transmitted into the first layer of coating, given by \cite{Dannenberg:09}

\begin{equation}
|\vec{E}_{\mathrm{surface}}| = |\vec{E_o^+}| |1 + \Gamma_1| .
\label{eq:2.4}
\end{equation}
$\vec{E_o^+}$ is set by the incident power density, which varies in different applications. Since $\Gamma_1$ has to be computed for evaluating the power reflectivity, the additional computational overhead is minimized. In order to minimize $\vec{E}_{\mathrm{surface}}$, and hence the absorption, it is conventional to add a nearly half-wavelength optical thickness `cap' of dielectric material to a high-reflectivity (HR) coating.

\subsection{Thermo-optic and Brownian noise}
\label{subsec:TOnoise}
Sensitivity of the current generation of laser interferometric gravitaitonal wave detectors is expected to be limited by the Brownian thermal noise of the coatings used. These consist of up to 20 pairs of alternating layers of $\mathrm{SiO}_2$ (low-index material) and $\mathrm{Ta}_2 \mathrm{O}_5$ (high-index material), or 40\,--\,50 pairs in materials with low index contrast (e.g. the pair of crystalline materials $\mathrm{AlGaAs}/\mathrm{GaAs}$~\cite{Cole:13}). In order to improve the sensitivity of future generation of detectors, an active area of research pursued in the last decade is the development of alternative dielectrics with which coatings that can meet the power reflectivity requirements can be developed.

Another application in which the Brownian noise of dielectric coatings can be a limiting noise source is the development of ultrastable frequency reference cavities~\cite{Robinson:19}. The frequency stability of lasers used in precision metrology experiments are often stabilized to such reference cavities, and efforts are underway to identify alternative dielectrics so that even more stable reference cavities can be constructed.

There is also a second effect which has to be simultaneously considered, the `thermo-optic' noise, which arises as a result of the temperature dependence of the refractive index of the dielectrics, and their thermally driven length fluctuations. With clever design, it is possible to suppress this noise contribution by coherent cancellation of the two effects~\cite{TaraC:16}.

\subsection{Immunity to small perturbations in assumed model parameters}
In evaluating the coating properties, assumptions are made about the layer thicknesses, refractive indices, dispersion, bulk absorption, and mechanical loss angle of the dielectrics. Additionally, uncertainties may exist, for example, in the assumed value of the angle of incidence. Since parameters of interest such as the power reflectivity of the coating are functions of these parameters, any uncertainty in them (due to manufacturing process limitations, measurement errors etc.) propagate through to the performance of the manufactured coating. In order to meet the tight tolerances on these parameters, it is desirable to choose (for fabrication) the coating design whose sensitivity to small errors in these model parameters is low.

We address this using a two step approach. First, during the optimization stage, the (numerical) derivative of coating properties with respect to model parameters are used in constructing the cost-function to be minimized. Secondly, the sensitivity of a given design to small ($\approx 1\%$) perturbations to the assumed model parameters is evaluated explicitly. The errors themselves are assumed to be i.i.d., and are drawn from an uncorrelated multivariate zero-mean Gaussian distribution whose standard deviation is chosen to be $0.5\%$ of the optimized value of the parameter. For example, errors in the thickness of the $i$-th layer, $l_i$, is sampled from the distribution $p(l_i) = \frac{1}{\sqrt{2 \pi \sigma^2}}e^{-l_i^2 / 2\sigma^2}$, with $\sigma = 0.005l_{i}^{\mathrm{opt}}$ and $l_{i}^{\mathrm{opt}}$ being the value of the thickness of the $i$-th layer that best achieves the design objectives.
The \texttt{emcee} package~\cite{emcee:13} is used to generate $\approx 1 \times 10^5$ samples, and the \texttt{corner} package~\cite{corner:16} provides a convenient way to visualize the results. A confidence interval on meeting specifications within tolerance can thus be stated. 

\section{Numerical optimization techniques for coating design}
\label{sec:psoMath}

\begin{figure}
    \centering
    \begin{subfigure}[b]{0.45\textwidth}
        \includegraphics[width=\textwidth]{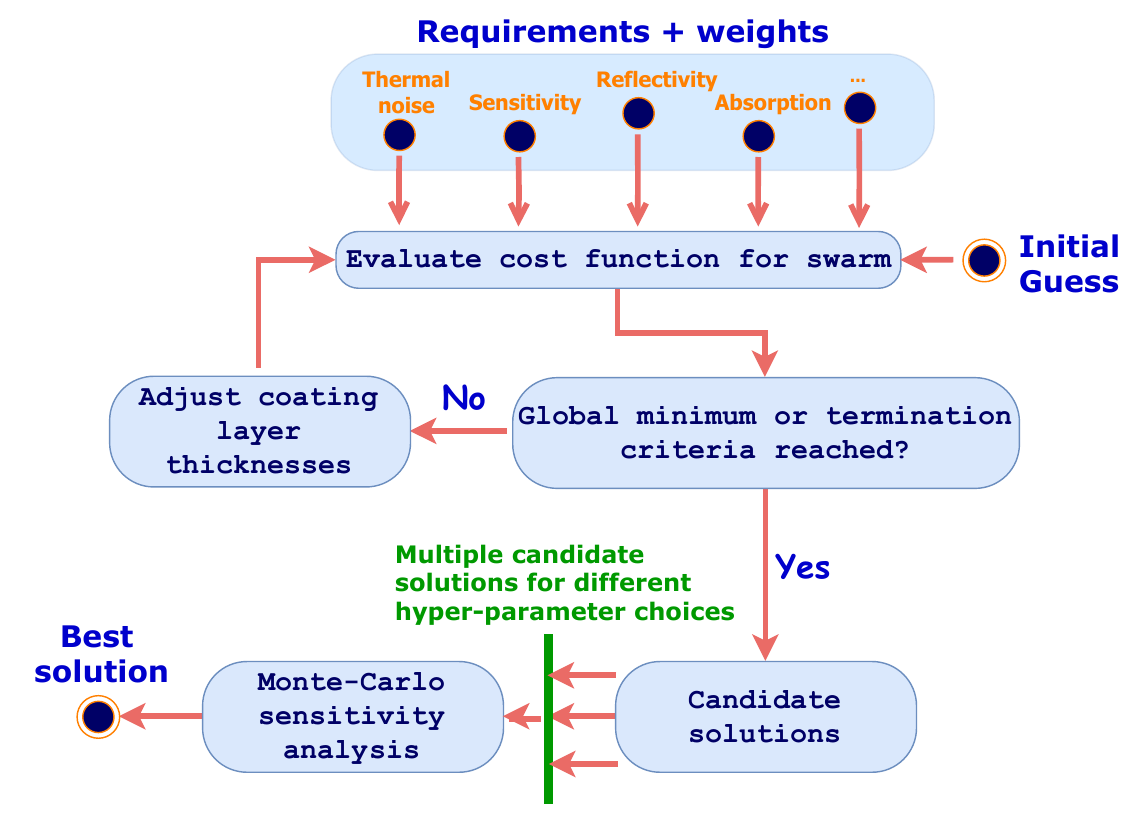}
        \caption{}
        \label{fig:3.1a}
    \end{subfigure}
	\quad
    \begin{subfigure}[b]{0.45\textwidth}
        \includegraphics[width=\textwidth]{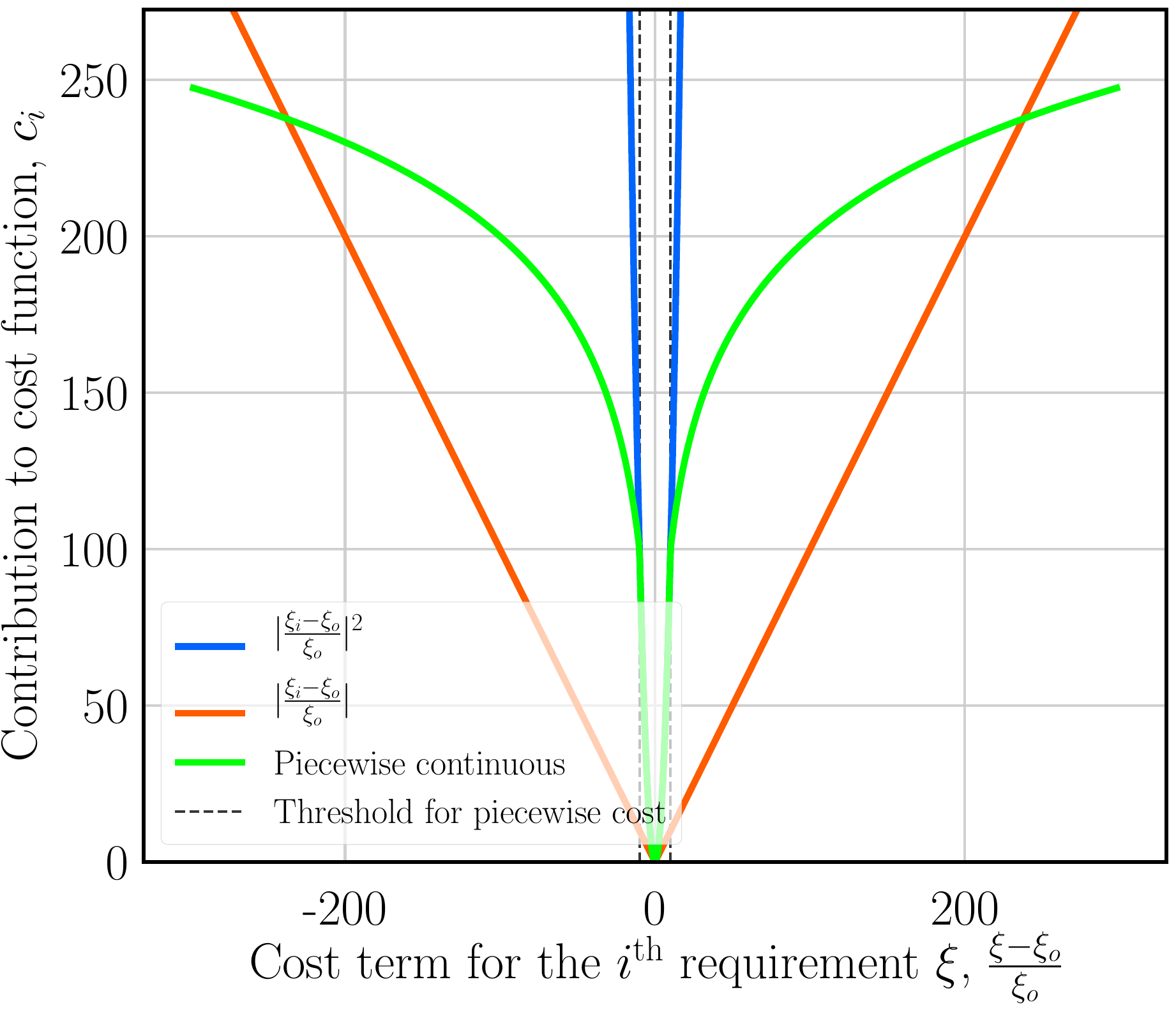}
        \caption{}
        \label{fig:3.1b}
    \end{subfigure}
    \caption{Overview of the workflow adopted. As shown in \Cref{fig:3.1a}, the process starts by identifying the requirements, and encoding these in a cost function. The cost function is then minimized using a global optimizer, such as \texttt{MATLAB}'s particle swarm optimization toolbox\G{\,\cite{MatlabPSO}} or \texttt{SciPy}'s  Differential Evolution\,\cite{2020SciPy-NMeth,Storn:97}. In order to choose between multiple candidate solutions, Monte-Carlo sensitivity analysis is used to choose the solution with the least sensitivity to model parameters. \Cref{fig:3.1b} shows the advantage of defining the cost function in a piecewise manner.}\label{fig:3.1}
\end{figure}

The methodology adopted for optimizing the dielectric coating design is schematically illustrated in \Cref{fig:3.1a}.

\subsection{Cost function construction and minimization}

The parameter space over which the optimization is done is the set of coating layer thicknesses, $\{ l_i \}$, and in the problems we consider, can be $\mathcal{O}(100)$-dimensional. Furthermore, the values $l_i$ can take are constrained to not be too small (for practical manufacturing reasons) and not larger than the longest optical wavelength of interest, $\lambda$. Once a set of requirements, $c_i$, and weights that reflect their relative importance, $w_i$ have been arrived at, these are encoded into a cost function, $\mathcal{C}(c_i, w_i, l_i)$. The weights are necessary to convert the \textit{multiple} objectives of the optimization problem into a \textit{scalar} which can be minimized using a numerical optimzation algorithm. The mathematical statement of the optimization problem amounts to
\begin{equation}
\begin{aligned}
& \underset{\{ l_i \}}{\mathrm{min}} & & \mathcal{C}(c_i, w_i, l_i), \\
& \text{subject to} & & l_{\mathrm{min}} \leq l_i \leq l_{\mathrm{max}} ~ \forall ~ i.
\end{aligned}
\label{eq:3.1}
\end{equation}

We chose \texttt{MATLAB}'s Particle Swarm Optimization (PSO) toolbox for implementing \Cref{eq:3.1}. Typical runtimes on a machine that can perform 45 GFlops is $\mathcal{O}(10)$ minutes. Subsequently, we also found similar performance could be realized using \texttt{SciPy}'s Differential Evolution optimizer\,\cite{2020SciPy-NMeth}, which has the added advantage of being an open-source utility. The functional form of the individual terms contributing to the cost function, $c_i$, was deliberately constructed in a normalized, piecewise manner. Normalization was necessary in order to compare costs from different requirements. Furthermore, since PSO looks for a globally optimal solution, it is likely that individual particles will traverse regions of high cost in the multi-dimensional parameter space. Defining the cost in a piecewise manner preserves a derivative term that allows the PSO functions to converge to a global minimum, but varied quadratically near the desired values and only logarithmically above some threshold value. As illustrated in \Cref{fig:3.1b}, this prevents the cost function from blowing up to large values for poor candidate solutions, as compared to other ways of specifying the error such as the $\mathrm{L}2-$ and $\mathrm{L}1-$ norms. Other functional forms, such as $\mathrm{sinh}^{-1} \left( |\frac{\xi - \xi_o}{\xi_o}| \right)$ can also be used to achieve the same goal.

\subsection{Monte-Carlo sensitivity analysis}
\label{subsec:MCHammer}
As there are "hyper-parameters" that do not directly enter the cost function, such as the number of layers in a coating, it is possible to have the optimization algorithm yield multiple candidate solutions. We wish to choose the simplest solution (from a fabrication standpoint) that is least sensitive to small perturbations in model parameters, and meets the coating requirements within specified tolerances. For the cases considered in this paper, the perturbations used are summarized in \Cref{tab:3.2.1}.

\subsubsection{Application to the inverse problem}
\label{sec:inverse}
\begin{figure}
    \centering
    \begin{subfigure}[b]{0.445\textwidth}
        \includegraphics[width=\textwidth]{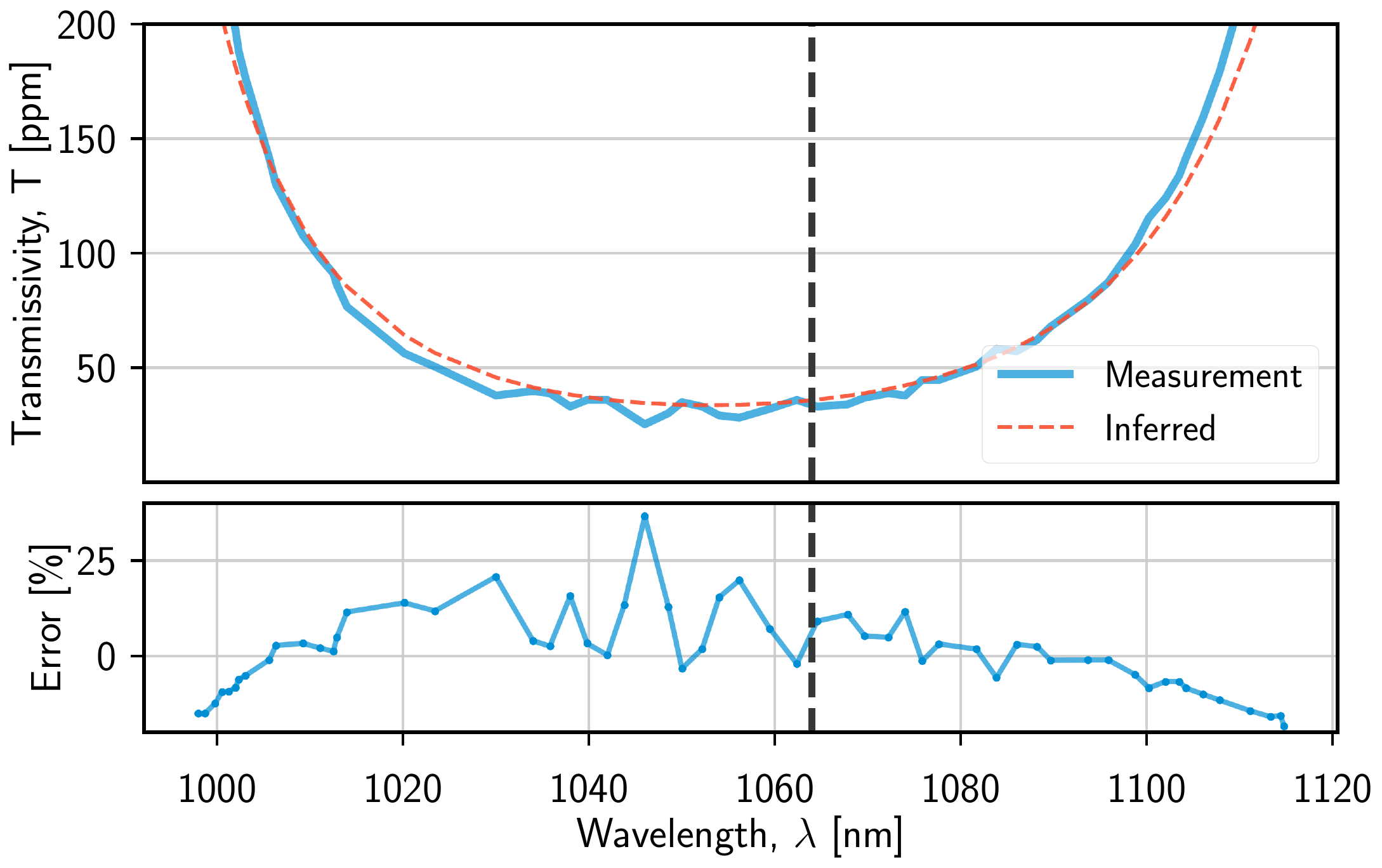}
        \caption{}
        \label{fig:3.2a}
    \end{subfigure}
	\quad
    \begin{subfigure}[b]{0.48\textwidth}
        \includegraphics[width=\textwidth]{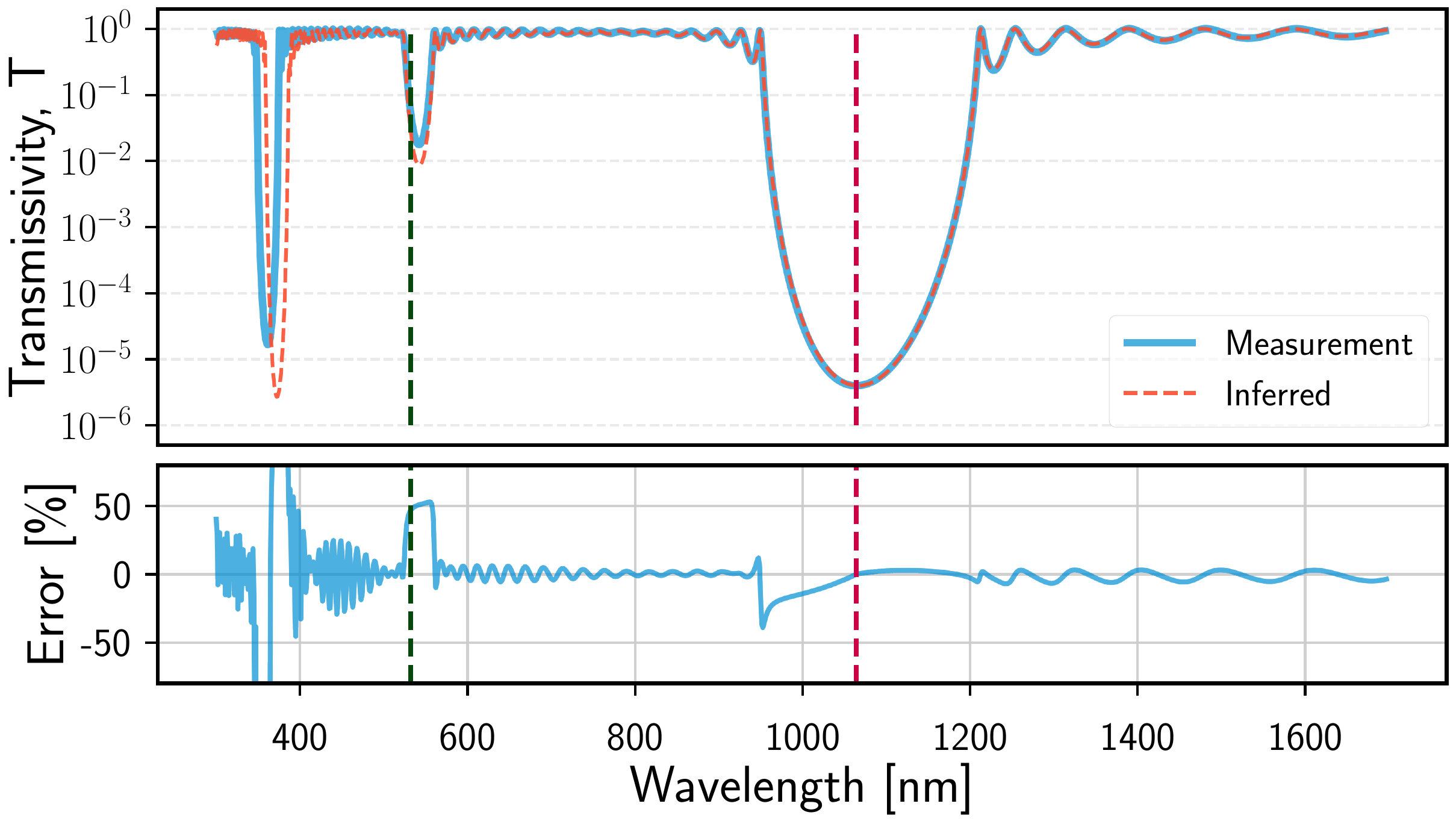}
        \caption{}
        \label{fig:3.2b}
    \end{subfigure}
    \caption{Inferring the coating structure from a spectral reflectivity measurement for the harmonic separator described in \Cref{subsec:HarmonicSeparator} (left), and the Advanced LIGO ETM (right). In both cases, critical wavelengths for the design are indicated by dashed vertical lines.}
\end{figure}

An interesting application of the Monte-Carlo approach is to apply it to the inverse problem of inferring the optical thicknesses of a dielectric coating, given a \emph{noisy measurement} of its spectral reflectivity as a function of the wavelength, $\lambda$. The problem becomes computationally expensive to evaluate if the dimensionality is unrestricted - that is, if we allow the physical thicknesses and refractive indices of all layers to be arbitrary. However, in practise, the dimensionality of the problem can be reduced. For concreteness, consider an optical coating composed of two dielectrics, $\mathrm{SiO}_2$ and $\mathrm{Ta}_2\mathrm{O_5}$. Assume the coating is built up with 19 repeated identical bilayer pairs, with the top bilayer pair having a different thickness for reducing the surface electric field amplitude. Furthermore, the dispersion of the dielectrics composing the coating are well characterized, and so may be taken as fixed. In this example, the task then amounts to the following - given the power transmissivity $T(\lambda)$, can we infer 4 numbers: $l_1$, the thickness of the top layer of $\mathrm{SiO}_2$, $l_2$, the thickness of the next layer of $\mathrm{Ta}_2\mathrm{O_5}$, and $[l_3, l_4]$, the thicknesses of the repeated bilayer pair? Applying this approach to the Harmonic Separator described in \Cref{subsec:HarmonicSeparator}, we infer thicknesses for the constructed coating that are within manufacturing tolerances. The modelled spectral reflectivity curve for the inferred coating is in good agreement with the measurement, as shown in \Cref{fig:3.2a}.

Having validated the technique, we applied it to a more complex problem, namely the Advanced LIGO ETM, results for which are shown in \Cref{fig:3.2b}. At shorter wavelengths, we found some deviation between the model constructed using inferred layer thicknesses, and the measured data - this was due to the unavailability of an accurate dispersion model for the dielectrics. Nevertheless, the technique yielded excellent results at the primary wavelength of interest, 1064\,nm. 

\begin{threeparttable}[htbp]
\centering
\caption{\bf Summary of perturbations (assumed to be uncorrelated, and hence, sampled from a multi-dimensional Gaussian probability distribution with diagonal covariance matrix). These were motivated by considering realistic uncertainties on these parameters.}
\begin{tabularx}{\linewidth}{Y|Y}
\toprule
\multicolumn{1}{c}{\bf Model parameter} & \multicolumn{1}{c}{\bf Uncertainty [\%]} \\
\hline
Physical thickness of layers, $\Delta l_i$ \tnote{$\dagger$} & 0.5 \\
Refractive indices of layers, $\Delta n_1$ and $\Delta n_2$ & 0.5 \\
Angle of incidence, $\Delta \theta_1$ \tnote{$\ddagger$} & 1 \\
\bottomrule
\end{tabularx}
\begin{tablenotes}
\item[$\dagger$] Due to the nature of the manufacturing process, all layers are expected to have identical \textit{fractional} uncertainties in \textit{physical} thickness.
\item[$\ddagger$] Applicable only for non-normal incidence cases.
\end{tablenotes}
\label{tab:3.2.1}
\end{threeparttable}

\section{Case studies and results}
\label{sec:results}
\Cref{tab:4.1} summarizes the requirements for a few case studies to which this methodology was applied. More details about the individual design requirements and results from the optimization runs are presented in the following subsections.

\begin{threeparttable}[htbp]
\centering
\caption{\bf Summary of the requirements on various parameters for three different coating designs that were optimized using the methodology outlined in this paper.}
\begin{tabularx}{\linewidth}{Y|ssss}
\toprule
\multicolumn{1}{c}{} & \multicolumn{3}{c}{\bf Case Studies} \\
\midrule
\bf Parameter & Harmonic separator& aLIGO ETM&$\mathrm{AlGaAs}$ ETM \tnote{*} & Voyager ETM\tnote{**}\\
\hline
Power reflectivity [\%] / transmissivity [ppm] \tnote{$\dagger$} & $T_{\lambda_1} \leq 50 $,  $R_{\lambda_2} \geq 99.9$ & $T_{\lambda_1} = 5 \pm 1$,  $R_{\lambda_2} = 97^{+2}_{-1}$ & $T_{\lambda_1} = 5 \pm 1$ & $T_{\lambda_1}\leq 6.5$\\
Thermo-optic noise [$\mathrm{m}/ \sqrt{\mathrm{Hz}}$] \tnote{$\ddagger$}& --- & $\leq 1.2 \times 10^{-21}$ & $\leq 1 \times 10^{-21}$ & $\leq2.3 \times 10^{-22}$ \\
Brownian noise [$\mathrm{m}/ \sqrt{\mathrm{Hz}}$]\tnote{$\ddagger$}& --- & $\leq 7.5 \times 10^{-21}$ & $\leq 2.5 \times 10^{-21}$ & $\leq5.0\times10^{-22}$ \\
Surface electric field [V/m] & --- & $\leq 1$ & $ \leq 2$ & $\leq 1$ \\
Absorption [ppm] & --- & $\leq 1$ & $\leq 1$ & $\simeq 3$~\cite{Steinlechner2017,Manel2023}\\
Angle of incidence [deg] & 41.1 & 0 & 0  & 0\\
Polarization \tnote{$\wedge$} & p-pol for $\lambda_1$, s- and p-pol for $\lambda_2$  & --- -
 & --- \\
\bottomrule
\end{tabularx}
\begin{tablenotes}
\item[*] $\mathrm{AlGaAs}$ is used to collectively refer to alternating layers of $\mathrm{Al}_{0.92}\mathrm{Ga}_{0.08}\mathrm{As}$ (low-index material) and $\mathrm{GaAs}$ (high-index material).
\item[$\dagger$] $\lambda_{1} = 1064 \mathrm{nm}$, $\lambda_{2} = 532 \mathrm{nm}$.
\item[$\ddagger$] For noise requirements, numbers quoted are \textit{amplitude} spectral densities at 100 Hz.
\item[$\wedge$] Both polarizations are degenerate for normal incidence, as is the case for the ETMs.
\item[**] $\lambda_1 = 2050\, \mu$m, with $\alpha$-Si:SiN bilayers on crystalline Silicon substrate at 124 K
\end{tablenotes}
\label{tab:4.1}
\end{threeparttable}

\subsection{Harmonic separator}
\label{subsec:HarmonicSeparator}
The aLIGO interferometers use multiple wavelengths of laser light to sense and control interferometric degrees of freedom of the suspended optical cavities \cite{Mullavey:12, Izumi:12}. In this case, the objective was to design a harmonic separator that allowed extraction of light at the second harmonic, $\lambda_2 = 532 \mathrm{nm}$, from a folded optical cavity, while preserving high reflectivity for the fundamental light field at $\lambda_1 = 1064 \mathrm{nm}$. Furthermore, since the expected angle of incidence on this optic was $\approx 41.1^o$, the design had to meet the $R$ and $T$ specifications for both s- and p-polarizaitons at $532 \mathrm{nm}$, while only p-polarization was of interest at $1064 \mathrm{nm}$.

\begin{figure}
    \centering
    \begin{subfigure}[c]{0.45\textwidth}
        \includegraphics[width=\textwidth]{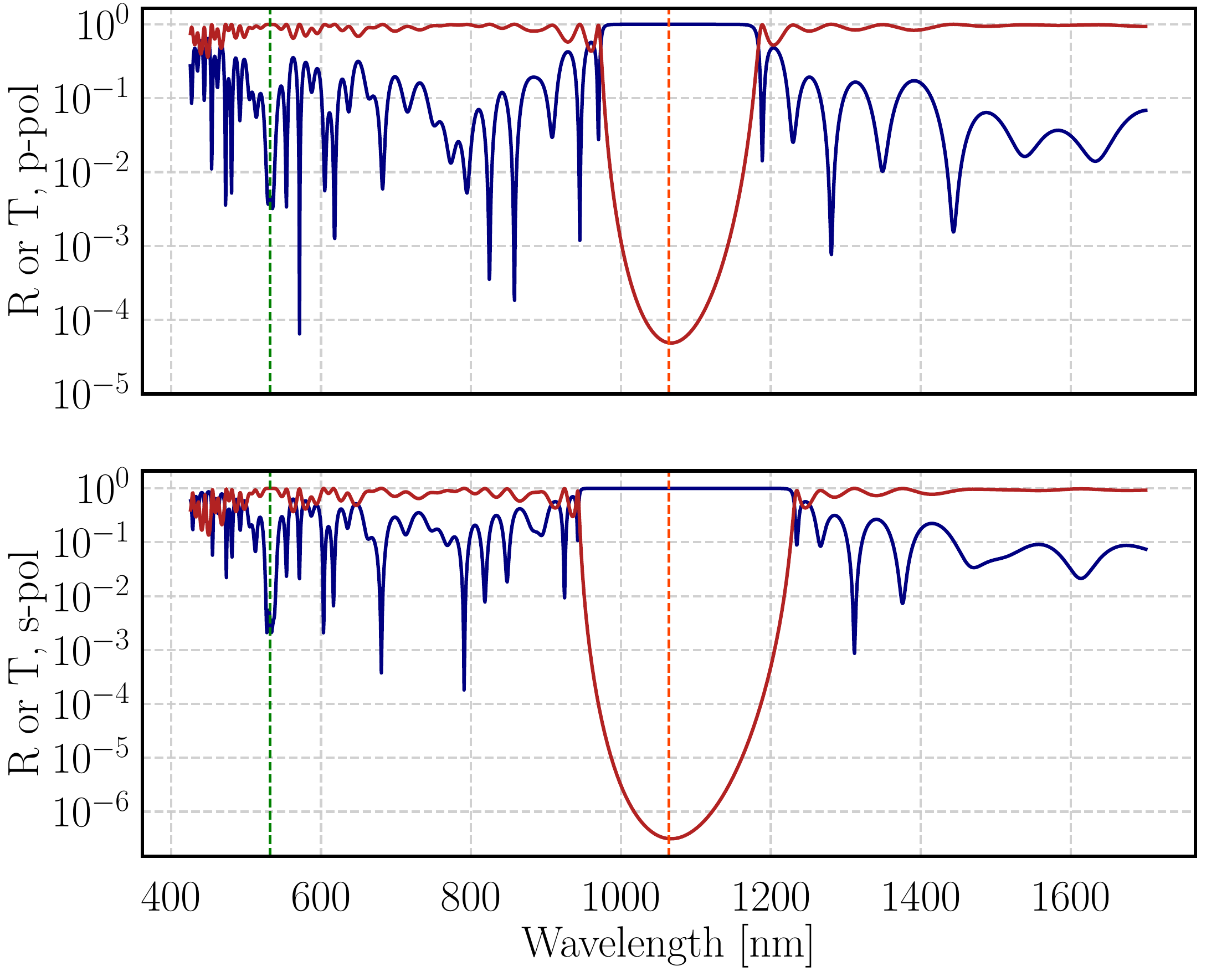}
        \caption{}
        \label{fig:4.1a}
        \includegraphics[width=\textwidth]{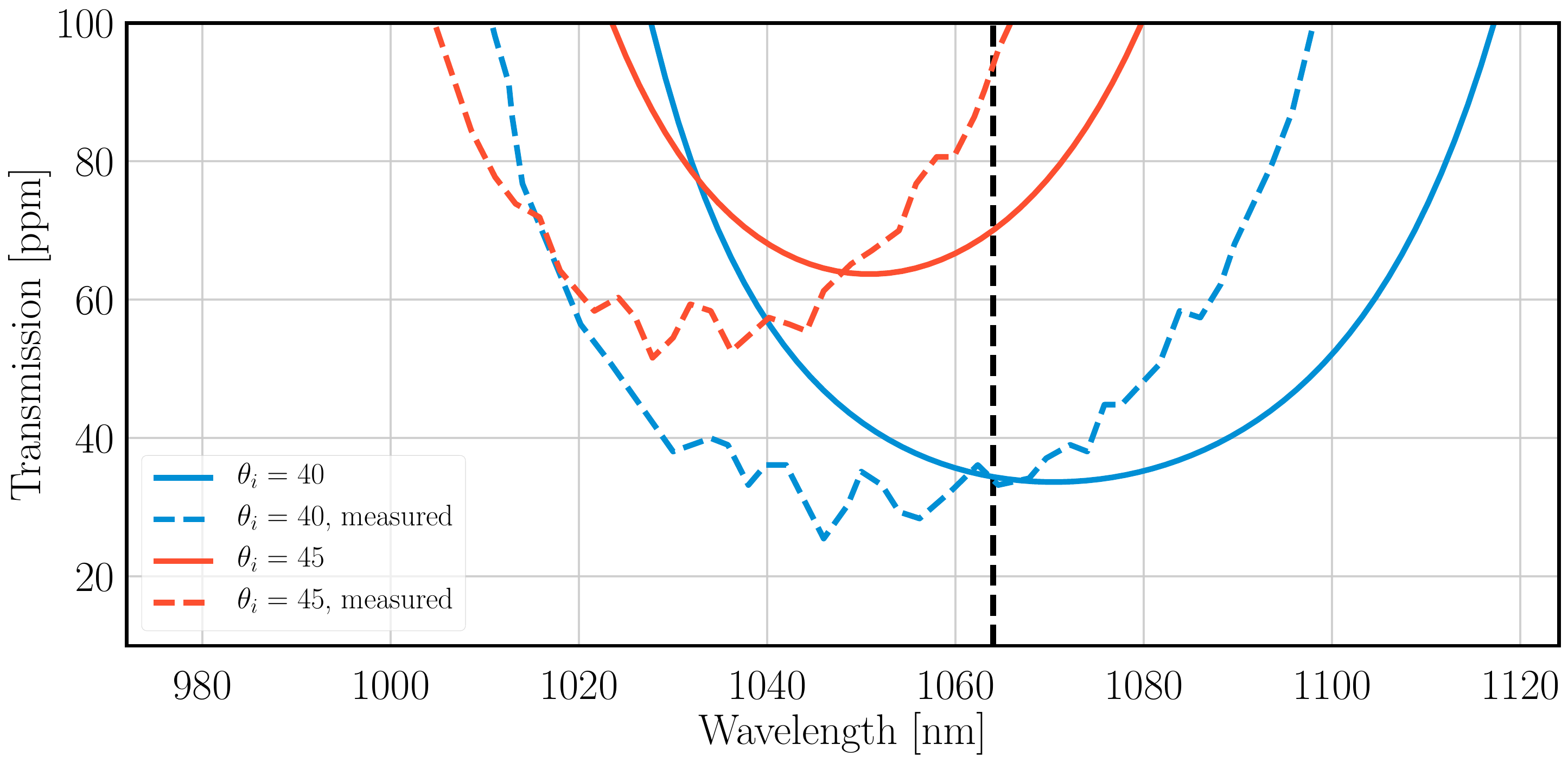}
        \caption{}
        \label{fig:4.1b}
    \end{subfigure}
    \quad
    \begin{subfigure}[c]{0.5\textwidth}
	\centering
        \includegraphics[width=\textwidth]{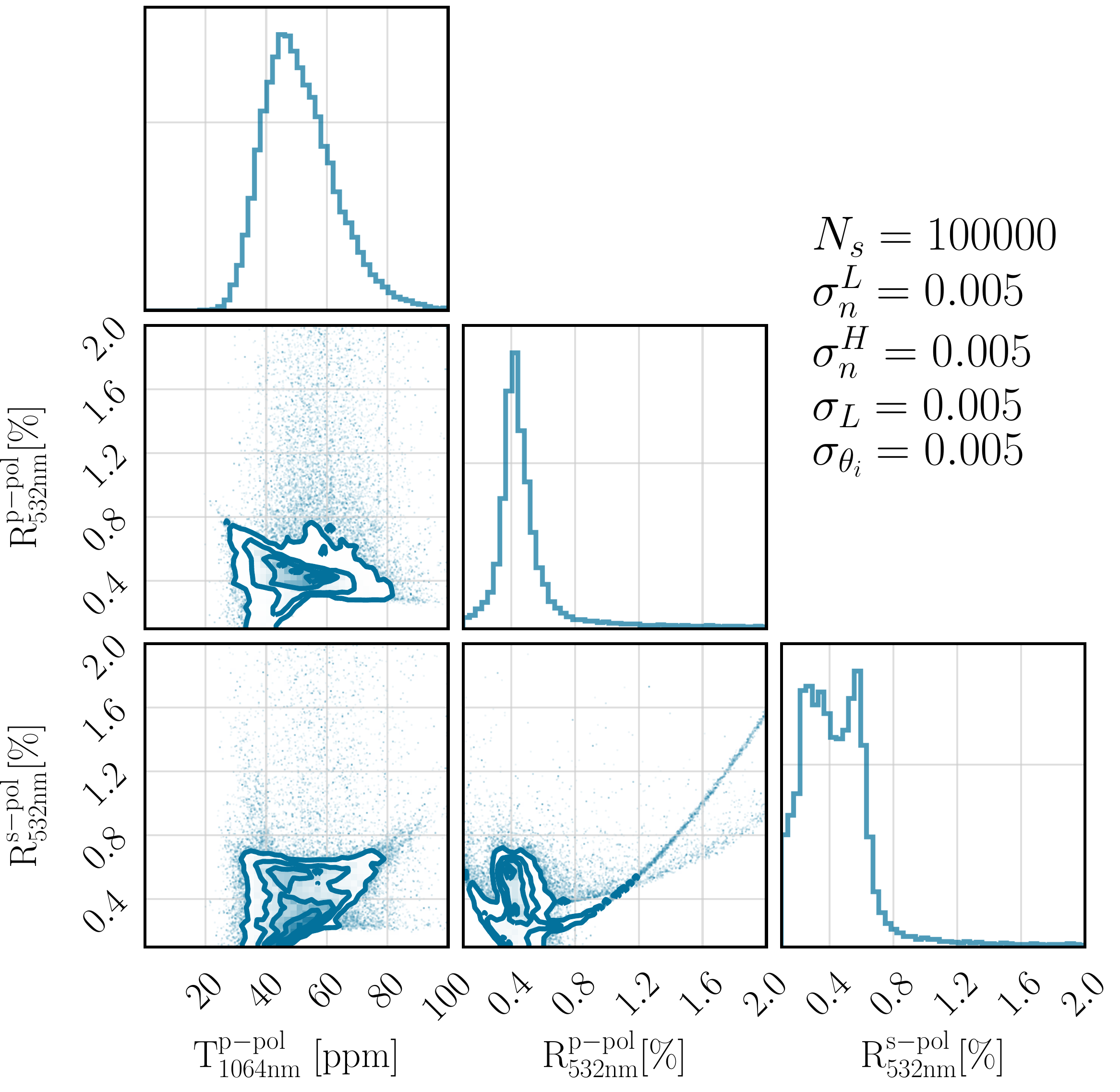}
        \caption{}
        \label{fig:4.1c}
    \end{subfigure}
    \caption{Performance of the optimized harmonic separator coating design. \Cref{fig:4.1a} indicates the wavelengths of interest, $\lambda_1$ (orange) and $\lambda_2$ (green), with dashed vertical lines. \Cref{fig:4.1b} shows the measured performance of a harmonic separator fabricated with layer thicknesses generated using the optimization routine described in the text. \Cref{fig:4.1c} shows the robustness of the design choice of 20 layer pairs in the dielectric stack to small variations in assumed model parameters.}\label{fig:4.1}
\end{figure}

 \Cref{fig:4.1a} shows the spectral reflectivity of the optimized coating design. \Cref{fig:4.1b} compares the \emph{measured} performance of a harmonic separator fabricated with layer thicknesses generated using this optimization routine. The measured spectral properties meet the design requirements, and are consistent with the tolerance analysis presented in \Cref{fig:4.1c}, which evaluates the sensitivity of the design to small perturbations, as described in \Cref{subsec:MCHammer}.

\subsection{HR cavity mirror coating for a gravitational wave detector}
The aLIGO interferometers are designed to have Fabry-P\'erot arm cavities with finesse $\approx 450$, for which the output coupling mirrors of the Fabry-P\'erot arm cavities (referred to as the End Test Mass, ETM) is required to have $T = 5 \pm 1 \mathrm{ppm}$ at $\lambda_1 = 1064 \mathrm{nm}$ \cite{ETMspec:09}. It is also required to have $R = 96.8\%$ at $\lambda_2 = 532 \mathrm{nm}$ to facilitate sensing and control of the arm cavity length using an auxiliary laser wavelength during the lock-acquisiton process \cite{Mullavey:12, Izumi:12}.

Additionally, since thermally driven microscopic fluctuations in the coating's optical and physical thickness is expected to limit the sensitivity of the instrument in the $100 \mathrm{Hz} \text{-} 1 \mathrm{kHz}$ frequency band, the chosen coating design's thermo-optic (TO) and Brownian noise spectral densities should not exceed specified thresholds in this frequency range. For $\mathrm{SiO}_2 / \mathrm{Ta}_2 \mathrm{O}_5$ coatings of the type used on the aLIGO optics, the mechanical loss angle, $\phi_{\mathrm{dielectric}}$ is $\approx 17 \times$ larger for $\mathrm{Ta}_2 \mathrm{O}_5$ than for $\mathrm{SiO}_2$ \cite{Amato:2021}, and so the total thickness of the latter in a given coating design dominates the thermal noise contribution. Hence, the design should be optimized for minimum thermal noise, while still meeting other requirements.

Finally, the circulating power in these Fabry-P\'erot cavities is expected to be $\mathcal{O}(1 \mathrm{MW})$ during high power operation. The coating should be designed with a safety factor such that it is not damaged under these conditions. One possible damage mechanism is that residual particulate matter on the optic's surface gets burnt into the coating. In order to protect against this, the coating has to be optimized to have minimum surface electric field.

\begin{figure}
  \begin{subfigure}[c]{\textwidth}
    \centering
      \includegraphics[width=0.75\textwidth]{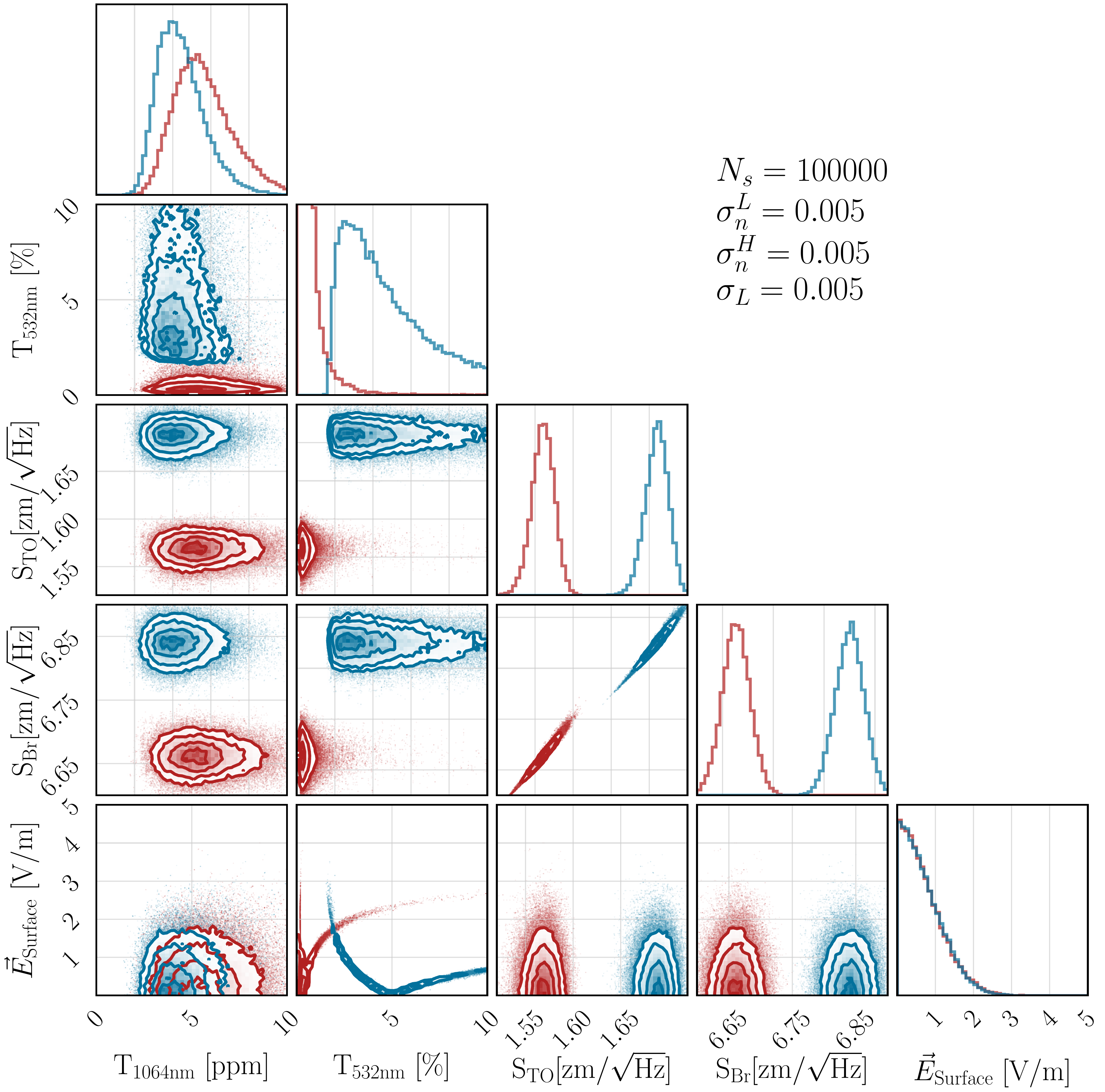}
  \end{subfigure}
  \caption{Performance of the optimized aLIGO ETM coating design. Here, we compare the robustness of the optimized design (red) and the as-built aLIGO ETM (blue), to small variations in assumed model parameters. The superior performance of our optimized coating, with respect to the design goals, is evident.
  }

  \label{fig:4.2}
\end{figure}

With these requirements as inputs to the optimization problem, we ran the particle swarm and sensitivity analysis and obtained a set of layer thicknesses. \Cref{fig:4.2} shows the performance of the optimized coating, and compares it to the as-built aLIGO ETM. The superior performance of our optimized design is evident - in particular, the transmissivity at $532\,\mathrm{nm}$ is much more robust in our design\footnote{In Advanced LIGO's first observing run, lock acquisition was made more difficult by the fact that the reflectivity of the ETM at $532\,\mathrm{nm}$ was $\approx 65\,\%$, while the design goal was $97\%$. \Cref{fig:4.2} shows that $T_{532\,\mathrm{nm}}$ for the as-built ETM coating design is indeed very sensitive to small errors in the model parameters.}. We assume $\phi_{\mathrm{Ti:Ta_2O_5}} = 3.6\times10^{-4}$ - this value is continually being revised by better measurements \cite{Amato:2021}, but since the same value is used to compare our design to the as-built aLIGO ETM, the conclusion that our design performs better remains valid.

\subsection{HR coatings for next-generation detectors}
As mentioned in \Cref{subsec:TOnoise}, alternative dielectric materials to $\mathrm{SiO}_2 / \mathrm{Ta}_2 \mathrm{O}_5$ are being considered in an effort to reduce the coating Brownian noise and improve the sensitivity of next-generation laser interferometric GW detectors. A promising alternative for wavelengths around 1 $\mu$m are alternating layers of low-index $\mathrm{Al}_{0.92}\mathrm{Ga}_{0.08}\mathrm{As}$ and high-index $\mathrm{GaAs}$ crystalline dielectrics shown to yield up to a 4-fold detector sensitivity improvement~\cite{Cole2023}. On the other hand, cryogenic Silicon detectors expected to operate at longer wavelengths~\cite{Voyager:15} have recently considered low-index $\mathrm{Si}_3 \mathrm{N}_4$ (silicon nitride) and high-index $\alpha$-$\mathrm{Si}$ (amorphous Silicon) to improve the coating thermal noise limit by over an order of magnitude relative to $\mathrm{SiO}_2 / \mathrm{Ta}_2 \mathrm{O}_5$ at 124\,K~\cite{Steinlechner2018,Steinlechner2017,Pan2018}.

\begin{figure}
    \centering
    \begin{subfigure}[c]{0.45\textwidth}
        \includegraphics[width=\textwidth]{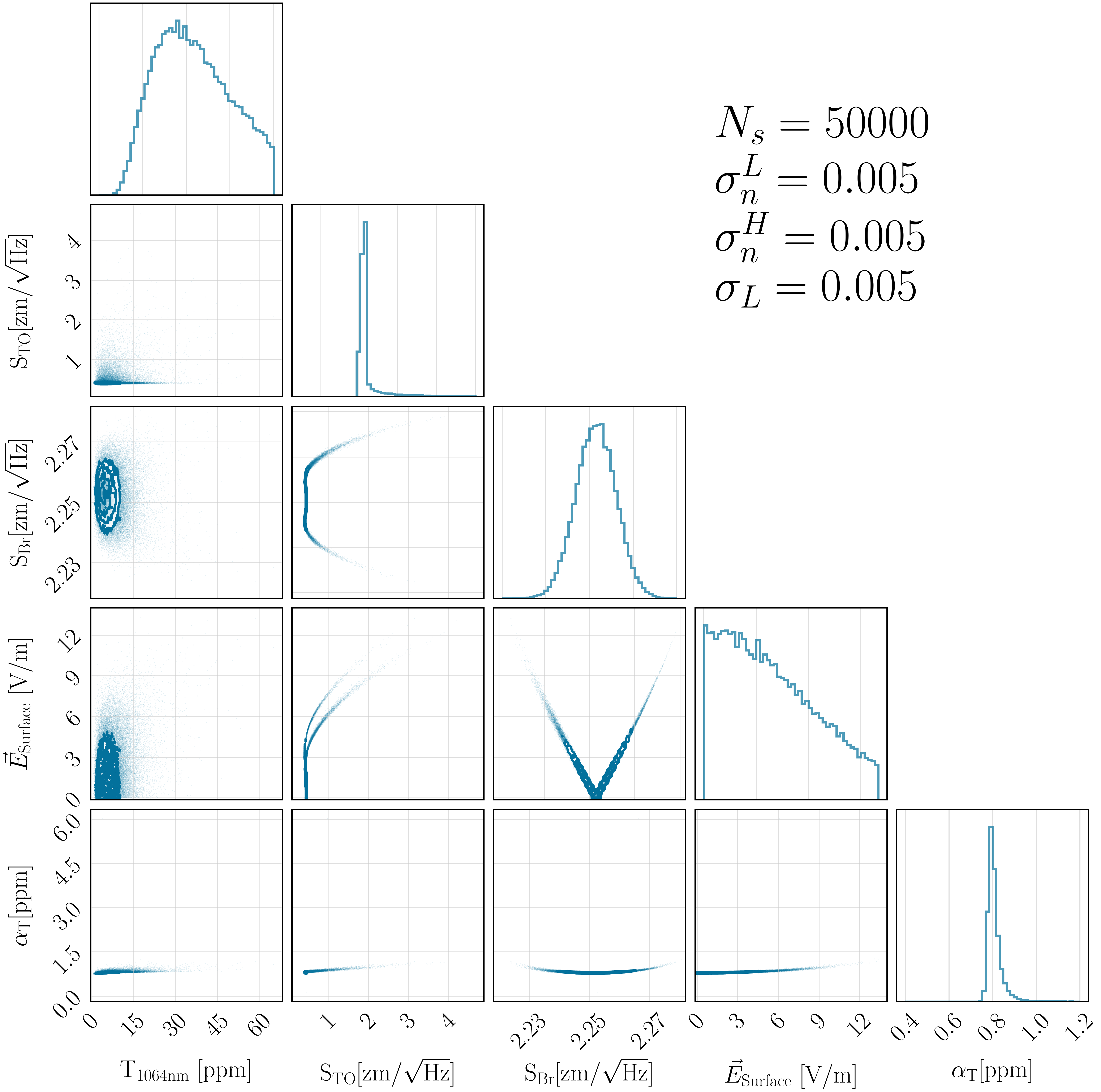}
        \caption{}
        \label{fig:4.3.1a}
    \end{subfigure}
    \quad
    \centering
    \begin{subfigure}[c]{0.45\textwidth}
        \includegraphics[width=\textwidth]{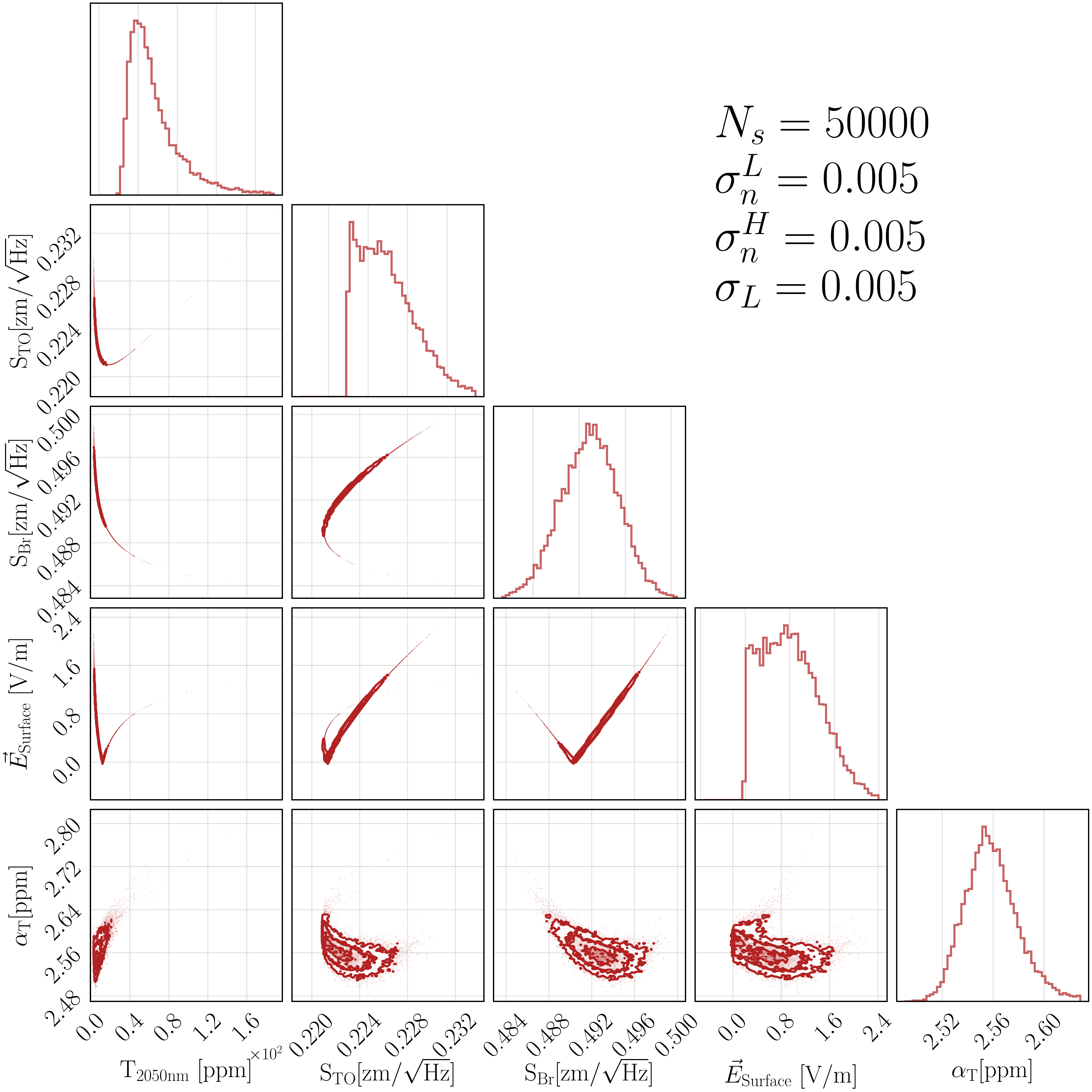}
        \caption{}
        \label{fig:4.3.1b}
    \end{subfigure}
    \begin{subfigure}[c]{0.45\textwidth}
	\centering
        \includegraphics[width=\textwidth]{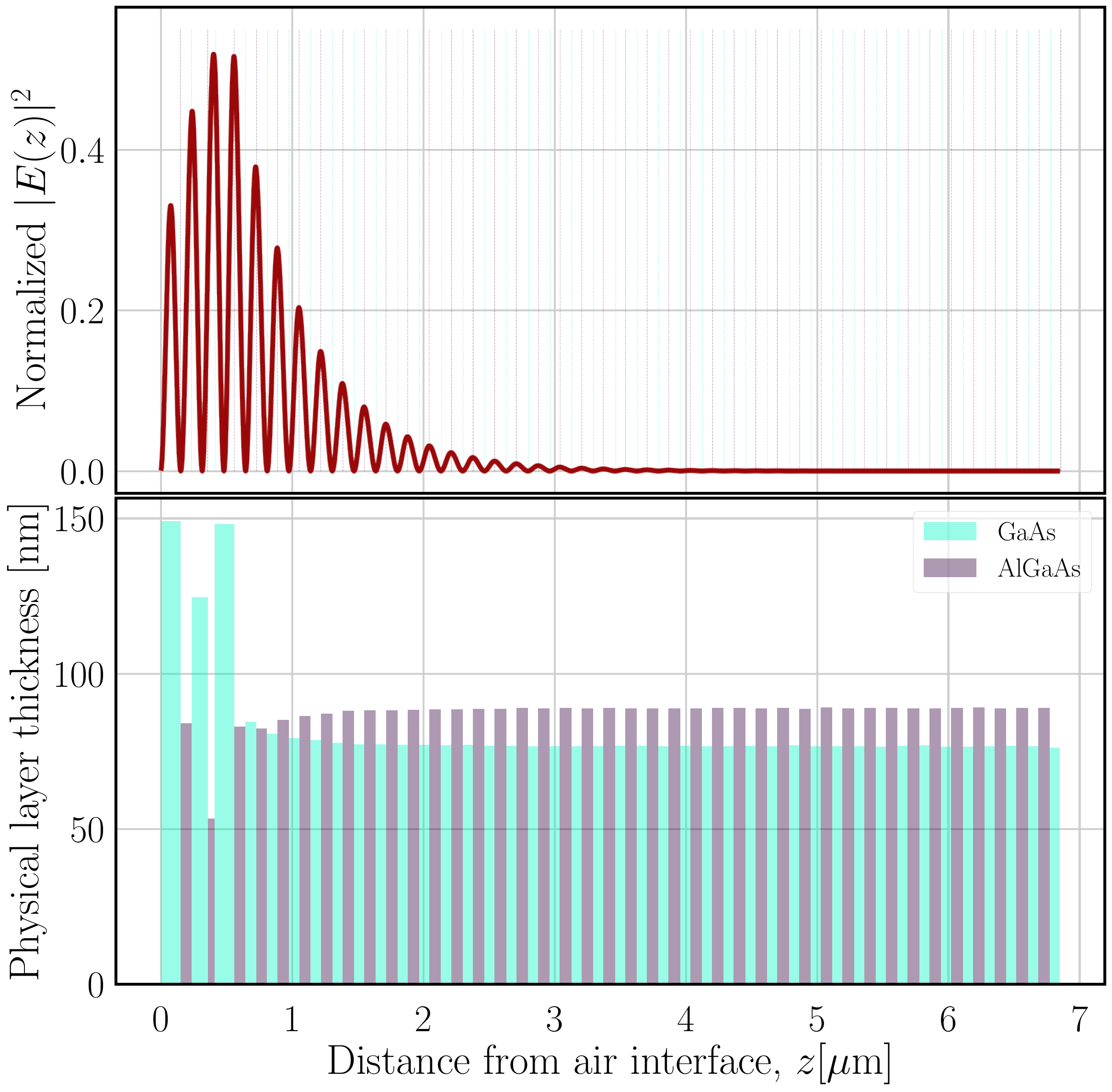}
        \caption{}
        \label{fig:4.3.1c}
    \end{subfigure}
    \quad
    \begin{subfigure}[c]{0.45\textwidth}
	\centering
        \includegraphics[width=\textwidth]{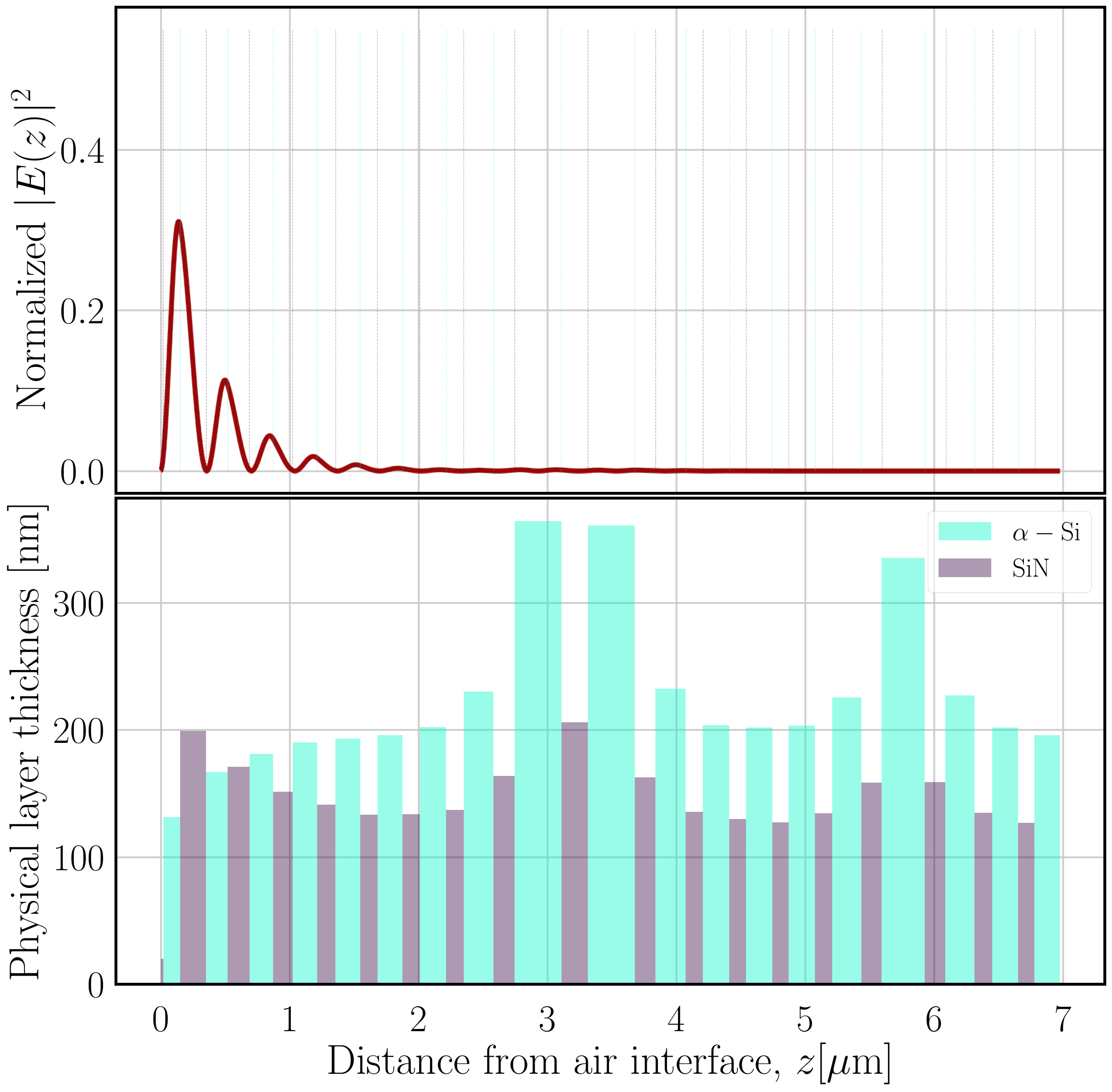}
        \caption{}
        \label{fig:4.3.1d}
    \end{subfigure}
    \caption{\Cref{fig:4.3.1a} and~\Cref{fig:4.3.1b} show the sensitivity analysis of the optimized AlGaAs:GaAs and $\alpha$Si:SiN coatings performance respectively. Below,~\Cref{fig:4.3.1c} and~\Cref{fig:4.3.1d} show the normalized square magnitude of electric field $|{\mathrm{E}(\mathrm{z})}/{\mathrm{E}_o^+}|^2$ inside the AlGaAs:GaAs and $\alpha$Si:SiN coatings respectively, along with the physical thickness profile of the individual dielectric layers after optimization. Dashed vertical lines indicate boundaries between low- and high-index materials.}
    \label{fig:4.3.1}
\end{figure}

While the coating Brownian noise may be significantly lower due to the low mechanical loss in these thin films, the overall thermal noise has to take into account both the Brownian and thermo-optic noise contributions. The latter can be minimized by coherent cancellation of thermorefractive and thermoelastic effects. This is done by including a penalty for the TO noise at a representative frequency (we choose 100\,Hz) in the cost function. While the absorption is not explicitly included in the cost function that is minimized by PSO, we include it in the MC sensitivity analysis, and confirm that the likelihood of it lying within the acceptable range of $\leq 1 \mathrm{ppm}$ is high, even if there are small deviations in assumed model parameters. The overall performance of the optimized crystalline coating is shown in \Cref{fig:4.3.1a}, while a similar sensitivity plot is shown in~\Cref{fig:4.3.1b} for the optimized cryogenic $\alpha$Si:SiN coating. ~\Cref{fig:4.3.1c} and~\Cref{fig:4.3.1d} show the variation of the electric field inside the crystalline and cryogenic coatings respectively, along with their layer thickness profile.

\section{Survey of past work, and future outlook}
There have been many proposed optimization algorithms for realizing dielectric coating thicknesses \cite{Lemarchand:14, Birge:11, Tikhonravov:07, Macleod:2010}.
The general algorithm is to define a cost function based on the design objectives, and then use some algorithm to find the global minimum value of this cost function in the allowed parameter space. Popular choices for global cost function minimization are the genetic algorithm, needle optimization, and swarm optimization algorithms \cite{Campbell:19}. Compared to these past works, our approach can easily handle these multiple design goals in a weighted manner. Moreover, our algorithm preferentially selects coating designs that are minimally sensitive to manufacturing tolerances, thereby decreasing the probability of fabricating a coating that does not meet the design specifications. Our optimization code has yielded comparable or superior coating designs relative to commercial coating design software, and since it is built using \texttt{Python}, does not incur expensive licensing fees. Development of a GUI will make the experience even smoother for a casual user, while more advanced problems of interest in the coating community can be set up easily given the modular nature of the algorithm.

During the writing of this manuscript, others in the thin-film community have developed complementary tools that can readily be integrated with the work presented here. In particular, \cite{Luce:22} has leveraged recent developments from the machine-learning community to dramatically reduce the computational time for evaluating the recursive relations described in \Cref{sec:2.1}. Adapting that software suite to the inverse problem described in \Cref{sec:inverse} could offer accurate inference of thin-film structures from spectral \emph{measurements} for even more complex coatings than the Advanced LIGO ETM. Designing anti-reflective (AR) coatings for transmitting carefully prepared quantum states of light with minimal loss in optical systems is another case-study we are currently investigating.

\section{Conclusion}
We have presented a method to find a globally optimum solution to multilayer dielectric coating design problems such that multiple competing objectives are satisfied. We only use standard libraries available with \texttt{MATLAB}'s PSO toolbox and \texttt{Python} for the implementation of this algorithm. PSO was chosen as the optimization algorithm given its success in similar problems. We found that using different optimization algorithms, such as Simulated Annealing, did not result in significant improvements in computational time. The performance of our optimized coatings are superior to conventional designs such as the "quarter-wave stack", and are robust to small perturbations in assumed model parameters. We have demonstrated the successful application of this technique to multiple design problems with varying requirements, including a harmonic separator, an HR coatings for aLIGO, and next-generation detectors using crystalline or cryogenic thin films. While the numerical efficiency of the algorithm may be improved, in its current iteration, our code runs in a reasonable amount of time on a modern multi-core machine capable of $\approx 50 \, \mathrm{GFlops}$. We are presently in the process of modifying the code to use \texttt{Python} for the global minimization step as well, so that the entire design process may be done using open-source software.




\section*{Acknowledgments}
We are grateful for valuable discussions with the Optics Working Group of the LIGO Scientific Collaboration. We thank Ramin Lalezari for valuable insights into mirror coating deposition constraints. FSC gracefully acknowledges support from the Barish--Weiss postdoctoral fellowship. 



\bibliography{main}

\begin{thebibliography}{10}
\newcommand{\enquote}[1]{``#1''}

\bibitem{Martynov:16}
D.~V. Martynov, E.~D. Hall, B.~P. Abbott, R.~Abbott, T.~D. Abbott
  \emph{et~al.}, \enquote{Sensitivity of the {Advanced LIGO} detectors at the
  beginning of gravitational wave astronomy,} {\protect\JournalTitle{Phys. Rev.
  D}} \textbf{93}, 112004 (2016).

\bibitem{MatlabPSO}
\enquote{{Particle Swarm, MATLAB Global Optimization Toolbox},}  (R2016b). The
  MathWorks, Natick, MA, USA.

\bibitem{2020SciPy-NMeth}
P.~Virtanen, R.~Gommers, T.~E. Oliphant, M.~Haberland, T.~Reddy \emph{et~al.},
  \enquote{{{SciPy} 1.0: Fundamental Algorithms for Scientific Computing in
  Python},} {\protect\JournalTitle{Nature Methods}} \textbf{17}, 261--272
  (2020).

\bibitem{Storn:97}
R.~Storn and K.~Price, \enquote{Differential evolution - a simple and efficient
  heuristic for global optimization over continuous spaces,}
  {\protect\JournalTitle{Journal of Global Optimization}} \textbf{11}, 341--359
  (1997).

\bibitem{Orfanidis:2016}
S.~Orfanidis, \emph{Electromagnetic Waves and Antennas} (S.J. Orfanidis, 2016).

\bibitem{Voyager:Science}
R.~X. {Adhikari}, P.~{Ajith}, Y.~{Chen}, J.~A. {Clark}, V.~{Dergachev}
  \emph{et~al.}, \enquote{{Astrophysical science metrics for next-generation
  gravitational-wave detectors},} {\protect\JournalTitle{Classical and Quantum
  Gravity}} \textbf{36}, 245010 (2019).

\bibitem{Arnon:80}
O.~Arnon and P.~Baumeister, \enquote{Electric field distribution and the
  reduction of laser damage in multilayers,} {\protect\JournalTitle{Appl.
  Opt.}} \textbf{19}, 1853--1855 (1980).

\bibitem{Dannenberg:09}
R.~Dannenberg, \enquote{{LMA} computation of surface {E}-field acheivable for
  {Advanced LIGO ITM},} Tech. Rep. T0900626, {LIGO Laboratory, California
  Institute of Technology} (2009).

\bibitem{Cole:13}
G.~D. Cole, W.~Zhang, M.~J. Martin, J.~Ye, and M.~Aspelmeyer, \enquote{Tenfold
  reduction of brownian noise in high-reflectivity optical coatings,}
  {\protect\JournalTitle{Nature Photonics}} \textbf{7}, 644 EP -- (2013).
  Article.

\bibitem{Robinson:19}
J.~M. {Robinson}, E.~{Oelker}, W.~R. {Milner}, W.~{Zhang}, T.~{Legero}
  \emph{et~al.}, \enquote{{Crystalline optical cavity at 4 K with
  thermal-noise-limited instability and ultralow drift},}
  {\protect\JournalTitle{Optica}} \textbf{6}, 240 (2019).

\bibitem{TaraC:16}
T.~Chalermsongsak, E.~D. Hall, G.~D. Cole, D.~Follman, F.~Seifert
  \emph{et~al.}, \enquote{Coherent cancellation of photothermal noise in
  gaas/al 0.92 ga 0.08 as bragg mirrors,} {\protect\JournalTitle{Metrologia}}
  \textbf{53}, 860 (2016).

\bibitem{emcee:13}
D.~{Foreman-Mackey}, D.~W. {Hogg}, D.~{Lang}, and J.~{Goodman},
  \enquote{{emcee: The MCMC Hammer},} {\protect\JournalTitle{PASP}}
  \textbf{125}, 306--312 (2013).

\bibitem{corner:16}
D.~Foreman-Mackey, \enquote{corner.py: Scatterplot matrices in python,}
  {\protect\JournalTitle{The Journal of Open Source Software}} \textbf{24}
  (2016).

\bibitem{Steinlechner2017}
J.~Steinlechner, C.~Kr\"uger, I.~W. Martin, A.~Bell, J.~Hough \emph{et~al.},
  \enquote{Optical absorption of silicon nitride membranes at 1064 nm and at
  1550 nm,} {\protect\JournalTitle{Phys. Rev. D}} \textbf{96}, 022007 (2017).

\bibitem{Manel2023}
\enquote{Personal communication with manel molina ruiz,}  (2023).

\bibitem{Mullavey:12}
A.~J. Mullavey, B.~J.~J. Slagmolen, J.~Miller, M.~Evans, P.~Fritschel
  \emph{et~al.}, \enquote{Arm-length stabilisation for interferometric
  gravitational-wave detectors using frequency-doubled auxiliary lasers,}
  {\protect\JournalTitle{Opt. Express}} \textbf{20}, 81--89 (2012).

\bibitem{Izumi:12}
K.~Izumi, K.~Arai, B.~Barr, J.~Betzwieser, A.~Brooks \emph{et~al.},
  \enquote{Multicolor cavity metrology,} {\protect\JournalTitle{J. Opt. Soc.
  Am. A}} \textbf{29}, 2092--2103 (2012).

\bibitem{ETMspec:09}
R.~Dannenberg, \enquote{{Advanced LIGO End Test Mass (ETM) Coating
  Specification},} Tech. Rep. E0900068, {LIGO Laboratory, California Institute
  of Technology} (2009).

\bibitem{Amato:2021}
A.~Amato, G.~Cagnoli, M.~Granata, B.~Sassolas, J.~Degallaix \emph{et~al.},
  \enquote{{Optical and mechanical properties of ion-beam-sputtered
  ${\mathrm{Nb}}_{2}{\mathrm{O}}_{5}$ and
  ${\mathrm{TiO}}_{2}\text{\ensuremath{-}}{\mathrm{Nb}}_{2}{\mathrm{O}}_{5}$
  thin films for gravitational-wave interferometers and an improved measurement
  of coating thermal noise in Advanced LIGO},} {\protect\JournalTitle{Phys.
  Rev. D}} \textbf{103}, 072001 (2021).

\bibitem{Cole2023}
G.~D. Cole, S.~W. Ballmer, G.~Billingsley, S.~B. CataÃ±o~Lopez, M.~Fejer
  \emph{et~al.}, \enquote{{Substrate-transferred GaAs/AlGaAs crystalline
  coatings for gravitational-wave detectors},} {\protect\JournalTitle{Applied
  Physics Letters}} \textbf{122}, 110502 (2023).

\bibitem{Voyager:15}
{LIGO Scientific Collaboration}, \enquote{{Instrument Science White Paper},}
  Tech. rep., {LIGO Laboratory, California Institute of Technology} (2015).

\bibitem{Steinlechner2018}
J.~Steinlechner, I.~W. Martin, A.~S. Bell, J.~Hough, M.~Fletcher \emph{et~al.},
  \enquote{Silicon-based optical mirror coatings for ultrahigh precision
  metrology and sensing,} {\protect\JournalTitle{Phys. Rev. Lett.}}
  \textbf{120}, 263602 (2018).

\bibitem{Pan2018}
H.-W. Pan, L.-C. Kuo, L.-A. Chang, S.~Chao, I.~W. Martin \emph{et~al.},
  \enquote{Silicon nitride and silica quarter-wave stacks for low-thermal-noise
  mirror coatings,} {\protect\JournalTitle{Phys. Rev. D}} \textbf{98}, 102001
  (2018).

\bibitem{Lemarchand:14}
F.~Lemarchand, \enquote{Application of clustering global optimization to thin
  film design problems,} {\protect\JournalTitle{Opt. Express}} \textbf{22},
  5166--5176 (2014).

\bibitem{Birge:11}
J.~R. Birge, F.~X. K\"{a}rtner, and O.~Nohadani, \enquote{Improving thin-film
  manufacturing yield with robust optimization,} {\protect\JournalTitle{Appl.
  Opt.}} \textbf{50}, C36--C40 (2011).

\bibitem{Tikhonravov:07}
A.~Tikhonravov, M.~Trubetskov, and G.~W~DeBell, \enquote{Optical coating design
  approaches based on the needle optimization technique,}
  {\protect\JournalTitle{Applied optics}} \textbf{46}, 704--10 (2007).

\bibitem{Macleod:2010}
H.~A. Macleod, \emph{{Thin-Film Optical Filters}} (CRC Press, 2010), chap. {16.
  Other topics}, pp. 729--740, {Series in optics and optoelectronics}, {Fourth}
  ed.

\bibitem{Campbell:19}
S.~D. Campbell, D.~Sell, R.~P. Jenkins, E.~B. Whiting, J.~A. Fan \emph{et~al.},
  \enquote{Review of numerical optimization techniques for meta-device design,}
  {\protect\JournalTitle{Opt. Mater. Express}} \textbf{9}, 1842--1863 (2019).

\bibitem{Luce:22}
A.~Luce, A.~Mahdavi, F.~Marquardt, and H.~Wankerl, \enquote{Tmm-fast, a
  transfer matrix computation package for multilayer thin-film optimization:
  tutorial,} {\protect\JournalTitle{J. Opt. Soc. Am. A}} \textbf{39},
  1007--1013 (2022).

\end{thebibliography}

\end{document}